\documentclass[acmsmall]{acmart}

\usepackage{amsmath}
\usepackage{algorithmic}
\usepackage{array}
\usepackage[skip=1pt,labelfont=bf]{caption}
\usepackage{graphicx}
\usepackage{longtable}
\usepackage{hyperref}
\usepackage[ruled,vlined,lined,commentsnumbered]{algorithm2e}
\usepackage{amsfonts}
\usepackage{color, colortbl}
\usepackage{enumitem}
\usepackage{fancybox}
\usepackage{framed}
\usepackage{fontenc}
\usepackage{listings}
\usepackage{lscape}
\usepackage{makecell}
\usepackage{multirow}
\usepackage{pifont}
\usepackage{rotating}
\usepackage{setspace}
\usepackage{subfigure}
\usepackage{url}
\usepackage{soul}
\usepackage{textcomp}
\usepackage{xcolor}
\usepackage{wrapfig}
\usepackage{balance}
\usepackage{microtype}

\newcolumntype{L}[1]{>{\raggedright\let\newline\\\arraybackslash\hspace{0pt}}m{#1}}
\newcolumntype{C}[1]{>{\centering\let\newline\\\arraybackslash\hspace{0pt}}m{#1}}
\newcolumntype{R}[1]{>{\raggedleft\let\newline\\\arraybackslash\hspace{0pt}}m{#1}}

\newboolean{showcomments}
\setboolean{showcomments}{true}
\ifthenelse{\boolean{showcomments}}
 { \newcommand{\mynote}[2]{
      \fbox{\bfseries\sffamily\scriptsize#1}
        {\small$\blacktriangleright$\textsf{\emph{#2}}$\blacktriangleleft$}}}
        { \newcommand{\mynote}[2]{}}

\newcommand{\attacktool}{MIST}

\newcommand{\toolname}{CodeGPTSensor+\xspace}

\newcommand{\datasetname}{{HMCorp}\xspace}

\newcommand{\taskone}{LLM-generated code detection}

\definecolor{lightgray}{gray}{0.9}
\definecolor{graybg}{gray}{0.9}
\definecolor{grayframe}{gray}{0.5}

\newcommand{\intuition}[1]{
\cornersize{0.2}
\setlength{\fboxsep}{5pt}
\noindent
\Ovalbox{
\parbox{0.96\linewidth}{
    \em {#1}
}
}
}

\makeatletter
\DeclareRobustCommand\onedot{\futurelet\@let@token\@onedot}
\def\@onedot{\ifx\@let@token.\else.\null\fi\xspace}

\makeatother

\AtBeginDocument{%
  }

\setcopyright{acmlicensed}
\copyrightyear{2025}
\acmYear{2025}
\acmDOI{XXXXXXX.XXXXXXX}

\acmJournal{TOSEM}

\begin{document}

\title{Detecting LLM-generated Code with Subtle Modification by Adversarial Training}

\author{Xin Yin}
\affiliation{%
  \institution{The State Key Laboratory of Blockchain and Data Security, Zhejiang University}
  \city{Hangzhou}
  \country{China}
  }
\email{xyin@zju.edu.cn}

\author{Xinrui Li}
\affiliation{%
  \institution{The State Key Laboratory of Blockchain and Data Security, Zhejiang University}
  \city{Hangzhou}
  \country{China}
  }
\email{lixinrui@zju.edu.cn}

\author{Chao Ni}
\authornote{This is the corresponding author.\\Chao Ni is also with Hangzhou High-Tech Zone (Binjiang) Blockchain and Data Security Research Institute, Hangzhou, China.}
\affiliation{%
  \institution{The State Key Laboratory of Blockchain and Data Security, Zhejiang University}
  \city{Hangzhou}
  \country{China}
  }
\email{chaoni@zju.edu.cn}

\author{Xiaodan Xu}
\affiliation{%
  \institution{The State Key Laboratory of Blockchain and Data Security, Zhejiang University}
  \city{Hangzhou}
  \country{China}
  }
\email{xiaodanxu@zju.edu.cn}

\author{Xiaohu Yang}
\affiliation{%
  \institution{The State Key Laboratory of Blockchain and Data Security, Zhejiang University}
  \city{Hangzhou}
  \country{China}
  }
\email{yangxh@zju.edu.cn}


\begin{abstract}
With the rapid development of Large Language Models (LLMs), their powerful code-generation capabilities have been widely applied in tasks like code completion and automated development, demonstrating the value of improving coding efficiency. 
However, the extensive use of LLM-generated code also raises several new challenges.
On the one hand, issues such as the regulation of code provenance, copyright disputes, and code quality have become increasingly concerning. 
%
%
How to effectively detect LLM-generated code and ensure its compliant and responsible use has become a critical and urgent issue.
On the other hand, in practical applications, LLM-generated code is often subject to manual modifications, such as variable renaming or structural adjustments. 
%
%
Although some recent studies have proposed training-based and zero-shot methods for detecting LLM-generated code, these approaches show insufficient robustness when facing modified LLM-generated code, and there is a lack of an effective solution.

To address the real-world scenario where LLM-generated code may undergo minor modifications, we propose \toolname, an enhanced version of CodeGPTSensor, which employs adversarial training to improve robustness against input perturbations.
\toolname integrates an adversarial sample generation module, \textbf{M}ulti-objective \textbf{I}dentifier and \textbf{S}tructure \textbf{T}ransformation (MIST), which systematically generates both high-quality and representative adversarial samples. 
This module effectively enhances the model's resistance against diverse adversarial attacks.
Experimental results on the \datasetname{} dataset demonstrate that \toolname significantly improves detection accuracy on the adversarial test set while maintaining high accuracy on the original test set, showcasing superior robustness compared to CodeGPTSensor.
\end{abstract}




\maketitle

\section{Introduction}
As Large Language Models (LLMs) continue to evolve, there is an increasing amount of research focused on the application of LLMs in software engineering tasks~\cite{yin2024multitask,yin2024thinkrepair,xia2023keep}.
In particular, LLMs (e.g., ChatGPT~\cite{openaichatgpt} and CodeLlama~\cite{roziere2023code}) for automated code generation have emerged as promising tools to improve coding productivity and efficiency~\cite{xu2022systematic,nair2023generating,yin2024you}.
Despite the significant advantages, a sentiment analysis study reveals that fear is the predominant emotion people associate with the code generation capabilities of ChatGPT~\cite{feng2023investigating}.
Concerns have been expressed regarding the impact of LLMs on software engineering, the programming community, and education.
For instance, Liu et al.~\cite{liu2023refining} find that ChatGPT-generated code frequently encounters quality issues, including code style and maintainability challenges, incorrect outputs, compilation and runtime errors, and performance inefficiencies. 
Several studies indicate that LLMs often produce vulnerable code~\cite{khoury2023secure, liu2023no, copilotflaw}.
In response to concerns over unreliable content generated by ChatGPT that could undermine the long-standing trust within the community, Stack Overflow, a prominent online platform for developers, has prohibited contributions from ChatGPT~\cite{stackoverflowbangpt}.
Additionally, several studies in the education domain report that students have started using ChatGPT to complete class assignments or even to engage in academic dishonesty during exams~\cite{nietzel2023more, susnjak2022chatgpt}.

To address these concerns, researchers have proposed LLM-generated code detection models~\cite{gpt2detector,mitchell2023detectgpt, robertaqa,chatgptzero,copyleaks,writer,aitextclassifier} that can predict the likelihood of a given piece of code being generated by an LLM, thereby enabling the differentiation between LLM-generated code and human-written code. 
Despite demonstrating considerable detection capabilities, these models still face certain limitations.
In practical applications, LLM-generated code often undergoes human modifications, such as variable renaming, code refactoring, and other minor changes.
While these modifications typically do not alter the core logic, they are sufficient to confuse existing detection models, significantly reducing their detection accuracy.
Notably, even the SOTA model (i.e, CodeGPTSensor~\cite{xu2024distinguishing}) also demonstrates insufficient robustness when facing modified LLM-generated code.
This issue is particularly prominent in the education sector. 
As LLMs become increasingly prevalent in programming tasks, students may leverage their powerful capabilities to complete assignments or exams, further modifying the generated code to evade detection.
Such behavior not only challenges the fairness of academic evaluation systems but also imposes higher demands on educational institutions to ensure academic integrity.
Moreover, in developer communities and software engineering contexts, developers frequently modify LLM-generated code by replacing identifiers with custom names, partially refactoring the code, or altering comments. 
These modifications may obscure the code's origin, making it difficult for detection models to identify its LLM-generated nature. 
Consequently, potential risks associated with using LLM-generated code may remain unaddressed.
Therefore, investigating how to maintain the robustness of detection models in scenarios where generated code undergoes minor modifications is not only critical for improving the performance of LLM-generated code detection models but also essential for enabling their effective deployment in real-world applications.

In order to mitigate the threat posed by modified samples to detection models, adversarial training has gained significant attention as a robust defense mechanism~\cite{yang2022natural, tian2023code, zhou2024evolutionary, du2023extensive}. 
By incorporating adversarial samples into the training process, adversarial training enables models to learn the characteristics of such samples, thereby enhancing their resilience to input perturbations.
However, existing adversarial sample generation methods still have limitations.
On the one hand, existing methods primarily rely on identifier substitution strategies to generate adversarial samples~\cite{yang2022natural, zhou2024evolutionary}.
On the other hand, these methods often fail to achieve an optimal balance between attack success rate, semantic consistency, and perturbation magnitude~\cite{yang2022natural, tian2023code}, potentially resulting in adversarial samples that are either insufficiently challenging or overly divergent from the original samples.
Therefore, generating adversarial samples that are highly representative, highly adversarial, and semantically consistent, while effectively enhancing model robustness through adversarial training, remains a critical challenge in current research.

To address these limitations, we propose \toolname, which uses adversarial training techniques to further enhance the robustness of the detection model when facing minor code modifications.
Specifically, in adversarial training, we propose an adversarial sample generation module called  \textbf{M}ulti-objective \textbf{I}dentifier and \textbf{S}tructure \textbf{T}ransformation (MIST), which integrates identifier replacement and code structure transformation to simulate manual code modification. 
This module generates semantically consistent yet highly adversarial samples to improve the model's resistance to perturbations.
Additionally, through a multi-objective optimization framework, we can strike a balance between attack success rate, semantic consistency, and the degree of disturbance, thereby generating high-quality adversarial samples that are more representative and challenging.  
This approach would help the model better adapt to real-world application scenarios and effectively improve its robustness.

To investigate the effectiveness of \toolname on distinguishing LLM-generated code from human-written code, we use a large-scale benchmark dataset, HMCorp~\cite{xu2024distinguishing}, which includes 288,508 Python samples and 222,335 Java samples, each consisting of a pair of human-written code and LLM-generated code. 
Experimental results demonstrate that \toolname outperforms CodeGPTSensor on the Python and Java test sets from HMCorp, with an accuracy of 0.992 and 0.970, respectively. 
Furthermore, \toolname shows an accuracy improvement of 272.8\% and 190.8\% on the Python and Java adversarial test sets, respectively, demonstrating enhanced robustness in scenarios where the code has undergone minor modifications.
In brief, the main contributions of this paper are summarized as follows:
\begin{itemize}[leftmargin=*]
    \item We propose MIST, which employs identifier substitution strategies and code structure transformation strategies. 
    A multi-objective optimization framework is used to balance attack success rate, semantic consistency, and perturbation magnitude, thereby generating high-quality adversarial samples for adversarial training.
    
    \item We use the adversarial samples generated by the MIST for adversarial training, enabling the \toolname to learn the characteristics of adversarial samples and thereby improve its robustness against input perturbations.

    \item We conduct extensive evaluation: (1) assessing the effectiveness and robustness of the adversarially trained model versus the original model; (2) comparing the proposed multi-objective adversarial sample generation module with baselines on several metrics; (3) evaluating adversarial samples' impact on enhancing the detection models' robustness.

    \item To facilitate future research in LLM-generated code detection, we have open-sourced our dataset and source code~\cite{gitreplication}.
\end{itemize}
\label{sec:introduction}

\section{Motivation}

\subsection{Challenges \& Opportunities}
\label{sec:motivating_example}

\subsubsection{Limited robustness of LLM-generated code detection models}

\begin{figure}[htbp]
    \centering 
    \includegraphics[width=1\textwidth]{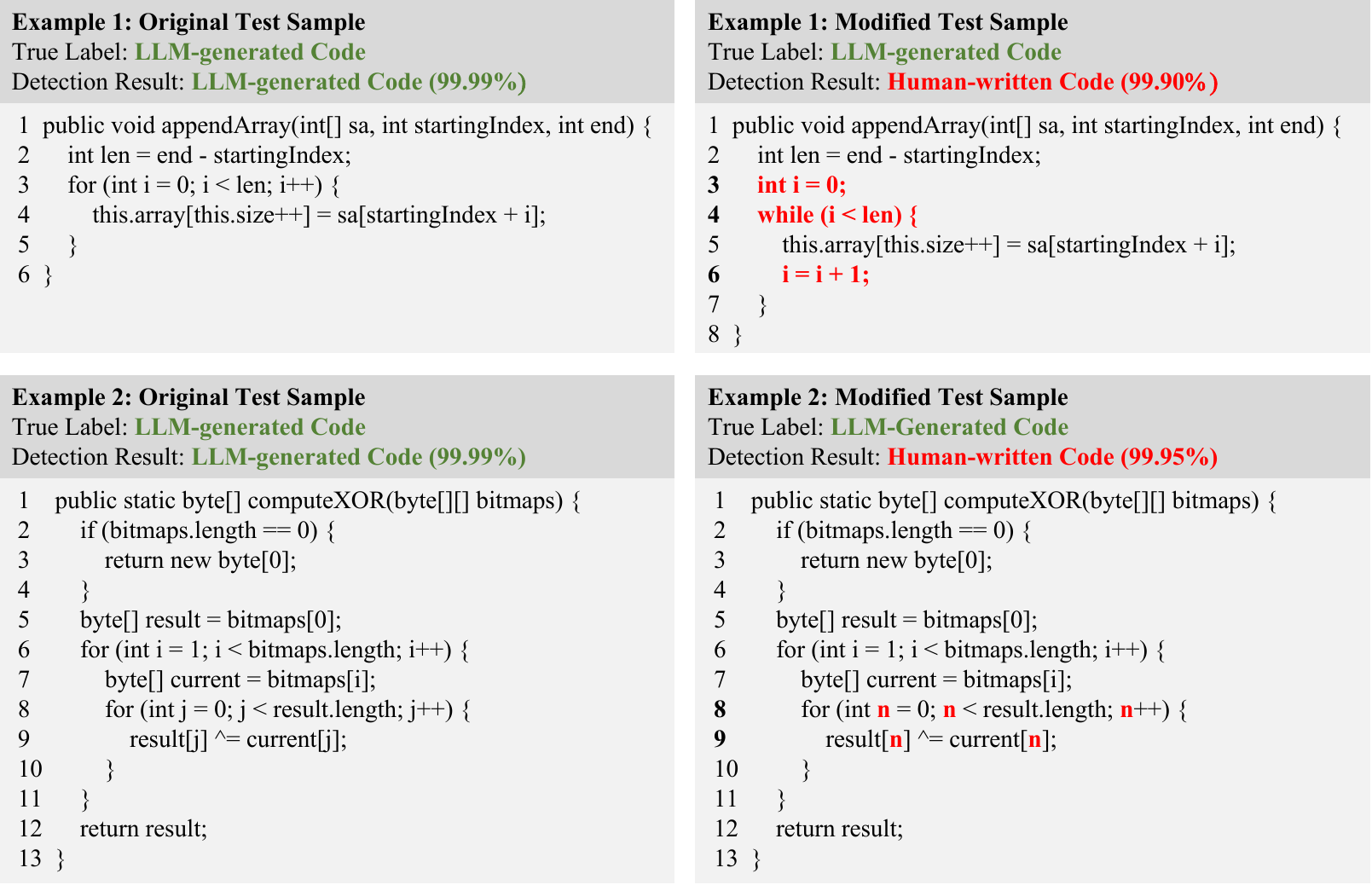}
    \caption{Detection results of CodeGPTSensor on original and modified test samples}
    \label{fig:motivation} 
\end{figure}

LLM-generated code detection models (e.g., CodeGPTSensor) have demonstrated excellent performance on various datasets (e.g., HMCorp). 
However, in practical applications, LLM-generated code is often subjected to minor human modifications, such as variable renaming or code structure adjustments.
While these modifications do not alter the core logic of the code, they can significantly impact the performance of existing detection models, leading to a sharp decline in detection accuracy. 
Taking CodeGPTSensor as an example, it performs excellently when detecting unmodified LLM-generated code, but it tends to fail when the code has undergone minor modifications. 
Figure~\ref{fig:motivation} shows two examples of minor modified code, where Example 1 changes a for loop to a while loop, and Example 2 replaces the variable name `j' with `n'.
The core logic remains consistent before and after the modification. 
However, CodeGPTSensor can accurately identify the original code as LLM-generated but mistakenly classifies the minor modified code as human-written with high confidence (i.e., 99.90\% in Example 1 and 99.95\% in Example 2).
This phenomenon reveals the limitations of current LLM-generated code detection models in terms of robustness when facing real-world scenarios.

\subsubsection{Limited effectiveness of adversarial example generation methods.}

In recent years, adversarial training has gained widespread attention to improve model robustness~\cite{yang2022natural, tian2023code, zhou2024evolutionary, du2023extensive}.
By incorporating adversarial examples during the training process, adversarial training enables models to learn the characteristics of these examples, thereby enhancing their robustness against input disturbances. 
In this process, the quality of adversarial examples is crucial. 
In the field of code, existing adversarial example generation methods are mainly divided into two categories: white-box attacks and black-box attacks.
White-box attacks rely on the internal information of the target model (e.g., gradients) to generate adversarial examples in a targeted manner, while black-box attacks only depend on the input-output interactions of the target model, making them more applicable in practical scenarios. 
However, existing black-box attack methods~\cite{yang2022natural, tian2023code, zhou2024evolutionary} have limitations when generating adversarial examples, which restrict the effectiveness of adversarial training.
On the one hand, most existing methods rely on identifier replacement strategies and fail to integrate code structure transformations effectively.
This makes it difficult to simulate the behavior of humans who often alternate between variable renaming and logic restructuring when rewriting code. 
This limits the diversity and representativeness of adversarial examples. 
On the other hand, existing methods struggle to achieve an ideal balance between attack success rate, semantic consistency, and the degree of disturbance, resulting in adversarial examples that are either not challenging enough or deviate too far from the original examples.

\subsection{Key Ideas}
Based on the above example analysis and technical background, we believe that adversarial training techniques can be leveraged to further enhance the robustness of CodeGPTSensor when facing minor code modifications. 
Specifically, during adversarial training, we can draw on human code rewriting behaviors by effectively combining identifier replacement and structural transformation strategies. 
Additionally, employing a multi-objective optimization framework allows us to balance the attack success rate, semantic consistency, and the degree of disturbance, thereby generating high-quality adversarial examples that are more representative and challenging.
This approach aids the model in better adapting to real-world scenarios and effectively enhancing its robustness.
\label{sec:motivation}

\section{Approach}
\label{sec:approach}

\begin{figure}[htb]
    \centering
    \includegraphics[width=\linewidth]{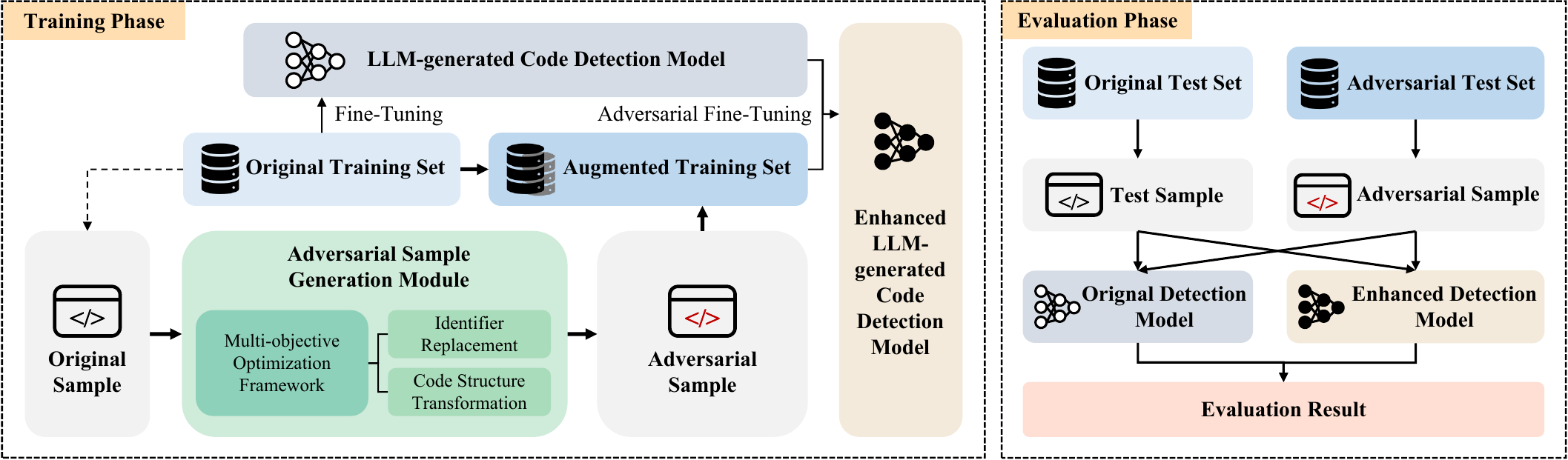}
    \caption{The framework of \toolname}
    \label{fig:framework}
\end{figure}

To address the issue of CodeGPTSensor's significant accuracy drop when detecting minor modified LLM-generated code, we propose an enhanced version, \toolname. 
\toolname incorporates an adversarial training strategy and designs an adversarial sample generation module called \textbf{M}ulti-objective \textbf{I}dentifier and \textbf{S}tructure \textbf{T}ransformation (MIST) to generate representative and high-quality adversarial samples.
This approach effectively improves the model's ability to adapt to input disturbances, thereby enhancing its robustness.
The overall framework of \toolname is shown in Figure~\ref{fig:framework}.
During the training phase, \toolname uses the MIST module to generate adversarial samples and mixes them with the original samples to construct an augmented training set. 
This augmented set is then used for adversarial training, enhancing the model's robustness against input disturbances. 
In the evaluation phase, the performance of the original model (i.e., CodeGPTSensor) and the enhanced model (i.e., \toolname) is assessed on both the original test set and the adversarial test set, providing a comprehensive evaluation of \toolname's performance in the LLM-generated code detection task.

\vspace{-0.3cm}
\subsection{Adversarial Sample Generation Module}
\label{sec:attack-module}
This section draws on the ideas of genetic algorithms to design an adversarial sample generation module called MIST, where the \textbf{Mutation} and \textbf{Crossover} operations are key components.
The mutation operation introduces small modifications to the original code, such as identifier replacement or structural adjustments, increasing the diversity of the code samples and expanding the search space. 
The crossover operation combines features from two parent samples to generate new offspring samples, helping to discover potential high-quality solutions within the existing population and accelerating the search process.
The overall process of the MIST module is shown in Algorithm~\ref{alg:attack}.
In lines 1 to 5, the population is initialized. 
The input code snippet $x$ undergoes $N$ mutations, and the resulting $N$ new samples form the initial population $\mathcal{P}$. 
Starting from line 6, the main loop begins.
First, the crossover is applied to the parent samples in the population, producing $N$ offspring, which are stored in $\mathcal{C}$. 
Starting from line 13, each sample in $\mathcal{C}$ undergoes mutation. 
Line 17 performs sample selection using the non-dominated sorting algorithm from NSGA-II~\cite{deb2000fast}, which selects the optimal $N$ samples based on multiple optimization objectives to form the next generation population $\mathcal{P}$, entering the next iteration.

\begin{algorithm}
    \begin{algorithmic}[1]
        \setstretch{1}  
        \REQUIRE input code $x$, population size $N$, max iteration \textit{max\_iter}
        \ENSURE population $\mathcal{P}$
        \STATE $\mathcal{P} = \emptyset$;
        \FOR {$i = 1$ to $N$}
            \STATE $x^i = \text{mutation}(x)$;
            \STATE $\mathcal{P} = \mathcal{P} \cup \{x^i\}$;
        \ENDFOR
        \WHILE{not exceed \textit{max\_iter}}
            \STATE $\mathcal{C} = \emptyset$;
            \FOR {$i = 1$ to $\frac{N}{2}$}
                \STATE Randomly select two parents $x^a$ and $x^b$ from $\mathcal{P}$;
                \STATE $\hat{x}^a, \hat{x}^b = \text{crossover}(x^a, x^b)$;
                \STATE $\mathcal{C} = \mathcal{C} \cup \{\hat{x}^a$, $\hat{x}^b\}$;
            \ENDFOR
            \FOR {$i = 1$ to $N$}
                \STATE $\hat{x}^i_{new} = \text{mutation}(\hat{x}^i)$;
                \STATE $C = (C\setminus\{\hat{x}^i\})\cup\{\hat{x}^i_{new}\}$;
            \ENDFOR
            \STATE $\mathcal{P} = \text{selection}(\mathcal{P} \cup \mathcal{C})$;
        \ENDWHILE
        \RETURN $\mathcal{P}$
    \end{algorithmic}
    \caption{The overall process of the MIST module}
    \label{alg:attack}
\end{algorithm}

In the above process, the specific steps of the mutation operation are shown in Algorithm~\ref{alg:attack-mutation}. 
The operation replaces identifiers with a probability of $r$, and with a probability of $1-r$, it performs a code structure transformation. 
The crossover operation is applied only to identifier replacement. 
For example, suppose the original input $x$ contains three variable names, $v1$, $v2$, and $v3$. 
The identifier replacement genes of two randomly selected parents $x^a$ and $x^b$ from $\mathcal{P}$ are represented as $<v1:a1, v2:a2, v3:a3>$ and $<v1:b1, v2:b2, v3:b3>$, respectively (where $v1:a1$ indicates replacing the identifier name $v1$ with $a1$). 
A random crossover point $h=2$ is chosen. 
Then, the offspring $\hat{x}^a$ and $\hat{x}^b$ after crossover have the genes $<v1:a1, v2:a2, v3:b3>$ and $<v1:b1, v2:b2, v3:a3>$, respectively.

\begin{algorithm}
    \begin{algorithmic}[1]
        \setstretch{1}  
        \REQUIRE input code $x$, identifier renaming rate $r$
        \ENSURE new code after mutation $x_{new}$
        \STATE $p \sim \mathcal{U}(0, 1)$;
        \IF{$p < r$}
            \STATE $x_{new} = \text{rename\_identifier}(x)$;
        \ELSE
            \STATE $x_{new} = \text{transform\_structure}(x)$;
        \ENDIF
        \RETURN $x_{new}$
    \end{algorithmic}
    \caption{The mutation operation of the \attacktool{} module}
    \label{alg:attack-mutation}
\end{algorithm}

Next, we will provide a detailed introduction to the identifier replacement strategy, code structure transformation strategy, and the definition of optimization objectives in \attacktool{}.

\subsubsection{Identifier Replacement Strategy}
\label{sec:identifier}
The identifier replacement strategy aims to introduce disturbances into code snippets by renaming identifiers. 
\attacktool{} uses \textbf{importance score (IS)} to measure the contribution of each identifier in the code snippet to the model's correct prediction. 
It also employs \textbf{context-aware identifier prediction} to ensure the naturalness of adversarial samples.

\textbf{(1) Identifier Importance Score: }When performing identifier replacement, the attack method first needs to determine which specific identifier to rename. 
Inspired by adversarial attack research in the fields of natural language processing and software engineering~\cite{zhou2024evolutionary, yang2022natural, garg2020bae}, \attacktool{} calculates the IS of each identifier in the code snippet to measure its contribution to the model's correct prediction. 
Specifically, for an input code snippet $x$, \attacktool{} first uses Tree-sitter~\cite{treesitter} to extract all the identifier names in the snippet and then calculates the IS for each identifier in turn.
When calculating the IS of the $i$-th identifier, the method replaces each occurrence of the identifier in the code snippet $x$ with the \texttt{<UNK>} mask, yielding a new code snippet denoted as $x^*_i$. 
The importance score of the $i$-th identifier for $x$ in the classification model $M$ is then computed as:

\begin{equation}
    \label{equ:is}
    IS_i=M(x)[y_\text{truth}]-M(x^*_i)[y_\text{truth}].
\end{equation}
\vspace{0.01cm}

In \autoref{equ:is}, $y_\text{truth}$ represents the true category of the code snippet $x$, and $M(x)[y_\text{truth}]$ represents the confidence that model $M$ assigns to the $y_\text{truth}$ category for $x$. 
Intuitively, $IS_i$ roughly represents the impact of the value of the $i$-th identifier on the model's prediction for $x$. 
If $IS_i > 0$, it indicates that the $i$-th identifier contributes to the correct prediction of $x$. 
The larger the IS, the greater the impact of the identifier on the prediction result. 
Therefore, changing the name of this identifier is more likely to alter the model's prediction, and it should be prioritized for replacement. 
Thus, when performing identifier replacement, the probability of selecting the $i$-th identifier is given by:

\begin{equation}
    \label{equ:is-prob}
    P(i) = \frac{IS_i}{\sum_{j=1}^{n} {IS_j}},
\end{equation}
\vspace{0.1cm}

where $n$ is the total number of identifiers in the code snippet $x$.

\textbf{(2) Context-Aware Identifier Prediction: }After selecting the identifiers to be replaced, the next step is to determine what the replacements should be. 
Inspired by the work of Zhou et al.~\cite{zhou2024evolutionary}, \attacktool{} adopts a context-aware identifier prediction strategy based on CodeT5~\cite{wang2021codet5} to generate more natural and semantically consistent adversarial samples.
Specifically, \attacktool{} first replaces each occurrence of the identifier to be replaced in the code snippet $x$ with the \texttt{<extra\_id\_0>} mask (a special token in the CodeT5 model). 
Then, using T5Tokenizer, the masked code snippet is transformed into a Token ID sequence that can be processed by the CodeT5 model. 
The CodeT5 model is used to predict the top-k candidate identifiers for the masked positions. Finally, one identifier is randomly selected from the candidate list to replace the original identifier, generating a new adversarial sample.
It is important to note that to ensure the quality of the generated adversarial samples, \attacktool{} ensures that the randomly selected new identifier names comply with the identifier naming conventions of the corresponding programming language and do not duplicate any existing identifier names in the current code snippet. 
If the selected identifier does not meet these conditions, \attacktool{} will randomly select another identifier until it satisfies the requirements.

\subsubsection{Code Structure Transformation Strategy}
In the \attacktool{} module, the code structure transformation strategy perturbs the code structure using a series of \textbf{equivalent transformation rules} to generate adversarial samples that maintain the same functionality but differ in syntax or structure. 
Additionally, \attacktool{} introduces a \textbf{code style-aware probabilistic transformation mechanism} to enhance the attack effectiveness of the adversarial samples.

\textbf{(1) Code Structure Equivalent Transformation Rules:}
\attacktool{} perturbs the code structure based on a series of equivalent transformation rules, ensuring that the transformed code maintains both syntactical correctness and semantic consistency. 
Inspired by existing research on code equivalence transformations~\cite{cheers2019spplagiarise, henkel2022semantic, tian2023code}, \attacktool{} employs three types of code structure equivalent transformation rules, as shown in \autoref{tab:structure-rule}.
These rules are applicable to commonly used programming languages such as Python and Java and include transformations for common constructs like loops, branches, increment/decrement operations, compound assignment operators, and equivalent replacements for constants and variables.
When applying the ''constant and variable equivalence conversion`` rule, \attacktool{} ensures that the temporary variable name used to replace the constant does not duplicate any existing variable names in the code snippet.
By applying these rules, slight perturbations to the code structure can be introduced without altering the program's functionality.
It is worth noting that, to ensure the naturalness of the generated adversarial samples, \attacktool{} does not use the commonly applied random insertion of dead code operations in code equivalence transformations~\cite{srikant2021generating, yefet2020adversarial}.

\begin{table}[htbp]
    \centering
    \resizebox{1\linewidth}{!}
    {
    \begin{tabular}{lp{30em}}
    \toprule
    \textbf{Category} & \textbf{Transformation Rules} \\
    \midrule
    Control Flow Equivalence Transformation &
    \makecell[l]{
    (1) Loop structure equivalence transformation, e.g., \texttt{for}  \texttt{while} \\
    (2) Branch structure equivalence transformation, e.g., \texttt{if-else} $\leftrightarrow$ \texttt{if-if}
    } \\
    \midrule
    Expression Equivalence Transformation &
    \makecell[l]{
    (3) Increment and decrement operator equivalence transformation, \\
    \hspace{2em}e.g., \texttt{j--} $\leftrightarrow$ \texttt{j=j-1} \\
    (4) Compound assignment operators \\
    \hspace{2em}(including \texttt{+=, -=, *=, /=, \%=, <<=, >>=, \&=, |=, \^}\texttt{=} \\ 
    \hspace{2em}and \texttt{>}\texttt{>}\texttt{>=} in Java, as well as \texttt{//=} and \texttt{**=} in Python) \\ 
    \hspace{2em}equivalence transformation, e.g., \texttt{x+=y} $\leftrightarrow$ \texttt{x=x+y}
    } \\
    \midrule
    Constant and Variable Equivalence Transformation & 
    \makecell[l]{
    (5) Constant and equivalent variable transformation, \\ 
    \hspace{2em}e.g., \texttt{print(``Hello, World!'')} $\leftrightarrow$ \\
    \hspace{2em}\texttt{message=``Hello, World!''} \\
    \hspace{2em}\texttt{print(message)} 
    } \\
    \bottomrule
    \end{tabular}%
    }
    \caption{Code Structure Equivalence Transformation Rules}
    \label{tab:structure-rule}
\end{table}%

\textbf{(2) Code Style-aware Probabilistic Transformation Mechanism: }After defining the code structure equivalent transformation rules, it is necessary to determine which transformation rules should be applied to a given code snippet. 
To achieve this, \attacktool{} designs a style-aware probabilistic transformation mechanism, allowing for more targeted structural perturbations and enhancing the effectiveness of the attack.
When performing adversarial attacks on the LLM-generated code detection model, \attacktool{} first extracts code style distribution information from the \datasetname{} training set. 
The specific steps are as follows:

\begin{itemize}[leftmargin=*]
\item \textbf{Divide Reference Samples: }The \datasetname{} is divided into subsets based on programming languages (Python and Java). 
Within each subset, the samples are further divided into human-written code and LLM-generated code.

\item \textbf{Statistical Structure Distribution Frequency: }For samples within each subset, the distribution frequency of different code structures is calculated.
Specifically, each rule in \autoref{tab:structure-rule} corresponds to two code structures, denoted as the original structure $S^b$ and the structure $S^a$ after applying the rule for transformation.
The occurrences of $S^b$ and $S^a$ in the sample set of each subset are counted as $N^b$ and $N^a$, respectively.

\item \textbf{Build Style Reference Table: }For each subset within the sample set of different sources, the relative frequency of code structures is calculated for each transformation rule: $\frac{N^a}{N^b+N^a}$ and $\frac{N^b}{N^b+N^a}$. These probability values form the style reference table.
\end{itemize}

When performing code structure transformations, \attacktool{} probabilistically applies the transformation rules based on the target sample's type and the style reference table. 
For example, suppose the target sample $x$ to be attacked has the true label ``LLM-generated Code'' and the programming language is Java. 
For each $S^b_i$ structure in $x$, under transformation rule $r_i$, the probability value $P^{Java}_{Human}(r_i) = \frac{N^a_i}{N^b_i + N^a_i}$ is obtained from the style reference table for the Java and human-written sample set.
The transformation rule is then applied with probability $P^{Java}_{Human}(r_i)$ to introduce structural perturbation, and with probability $1 - P^{Java}_{Human}(r_i)$, the original structure is preserved.
Through this probabilistic structural transformation, the structure distribution in the target sample gradually aligns with that of the reference samples.

\subsubsection{Definition of Optimization Objectives}
For the code classification model $M$, the goal of the adversarial attack is to apply slight perturbations to the code snippet $x$ that is correctly classified by $M$, thereby generating an adversarial sample $x^{\text{adv}}$ that causes the model to misclassify. 
Additionally, this adversarial sample should ensure that the code remains syntactically correct and the program's functionality does not change.
Beyond these basic requirements, researchers have proposed that adversarial code examples should also ensure ``naturalness'', meaning that, from a human perspective, the perturbations in the adversarial samples should appear natural~\cite{yang2022natural}. 
In other words, the replaced identifiers should be contextually appropriate, not arbitrary or unrelated names. 
Furthermore, the magnitude of the perturbation in the adversarial sample should not be excessive~\cite{zhou2024evolutionary}.

To balance multiple optimization objectives, such as attack success rate, semantic consistency, and perturbation degree, \attacktool{} employs a multi-objective optimization framework. 
This allows the generation of effective adversarial samples while maintaining semantic consistency and controllability of the modification degree.
Specifically, for a given code snippet $x \in \mathcal{X}$ with true label $y_{\text{truth}}$, the adversarial sample generated by perturbing $x$ is denoted as $x^{\text{adv}} \in \mathcal{X}$. The optimization objectives for generating adversarial samples for the classification model $M$ in \attacktool{} are as follows:

\begin{itemize}[leftmargin=*]
\item \textbf{Adversarial Loss (AL): }The confidence of model $M$ assigns to the true class $y_{\text{truth}}$ for $x^{\text{adv}}$:

\begin{equation}
    f_1(x^{\text{adv}})=M(x^{\text{adv}})[y_\text{truth}].
\end{equation}
\vspace{0.01cm}

\item \textbf{Semantic Distance (SD): }Calculated based on the cosine similarity of vectors:

\begin{equation}
    \label{equ:semantic-distance}
    f_2(x^{\text{adv}})=\sum_{i=1}^{n}1-\cos(v_i,v_i^{\text{adv}}),
\end{equation}
\vspace{0.04cm}

where $v_i$ is an identifier in $x$ and $n$ represents the number of identifiers in $x$.

\item \textbf{Edit Distance (ED): }Calculated based on the Levenshtein distance of the code:

\begin{equation}
    \label{equ:edit-distance}
    f_3(x^{\text{adv}})=\text{EditDistance}(x,x^{\text{adv}}).
\end{equation}

\end{itemize}

In the aforementioned optimization objectives, a smaller AL indicates that the model $M$ considers $x^{\text{adv}}$ less likely to belong to $y_{\text{truth}}$, implying a higher probability of a successful adversarial attack. 
A smaller SD signifies higher semantic similarity between the identifiers of $x$ and $x^{\text{adv}}$, ensuring more "natural" identifier substitutions. A smaller ED corresponds to fewer character-level edit operations required to transform $x$ into $x^{\text{adv}}$, thereby reducing perturbation magnitude. 
Consequently, the final multi-objective optimization of \attacktool{} can be formulated as:  

\begin{equation}  
\min_{\mathbf{x^{\text{adv}}}\in\mathcal{X}}\left(f_1(\mathbf{x^{\text{adv}}}),f_2(\mathbf{x^{\text{adv}}}),f_3(\mathbf{x^{\text{adv}}})\right).
\end{equation}
\vspace{0.03cm}

Notably, the formulation above omits explicit modeling of two critical constraints: ``syntactic correctness'' and ``functional equivalence''. 
This is because \attacktool{} employs two perturbation strategies—``identifier replacement strategy'' and ``code structure transformation strategy''—both of which are designed as syntax-preserving and functionality-invariant operations. 
These strategies guarantee that adversarial modifications strictly adhere to equivalent code transformations, thereby preserving the original code's grammatical validity and program behavior.

\subsection{Adversarial Training Strategy}
\toolname adopts the commonly used adversarial fine-tuning strategy in deep code model research~\cite{li2023comparative}. 
This strategy first trains the model on the original training set $D$ to obtain the optimal model $M$. 
Then, $M$ is used as the target model for adversarial attacks to generate an adversarial sample set $D^{adv}$. 
Next, the adversarial samples are added to the original training set to form an augmented training set $D^+$, which is used to fine-tune $M$, resulting in the enhanced model $M^+$.

Specifically, during the adversarial fine-tuning phase, in order to balance computational costs with the effectiveness of adversarial training and avoid excessive interference with the detection accuracy on the original dataset, \toolname randomly samples 10\% of the original training set as the target sample set for adversarial sample generation.
For each target sample in the set, \attacktool{} is used to perform an adversarial attack to generate an adversarial sample.
The first successful adversarial sample is selected if the attack is successful, meaning the target model's prediction changes. 
If the attack fails, the adversarial sample that reduces the target model's prediction probability for the correct class the most is chosen from the generated adversarial samples.
Subsequently, the samples from the original training set and the generated adversarial samples are mixed at a ratio of 70\% original samples and 30\% adversarial samples to construct the augmented training set. 
By using this augmented training set for fine-tuning, \toolname demonstrates stronger robustness against input disturbances while maintaining high detection accuracy on the original samples.
During fine-tuning, \toolname assumes that the true labels of adversarial samples are the same as those of the original samples and uses the cross-entropy loss, as shown in \autoref{equ:ce-loss}, to guide the model's optimization.

\begin{equation}
    \label{equ:ce-loss}
    \mathcal{L}_{ce} = -\sum_{i} y_i \log(\hat{y}_i).
\end{equation}

\section{Experiment Setup}
\label{sec:design}

This section introduces the studied dataset, the selected baselines, the evaluation metrics, and the implementation details.

\subsection{Studied Dataset}
\label{sec:dataset}

To evaluate the effectiveness of \toolname, we conduct experiments using the \datasetname{} dataset, consistent with the LLM-generated code detection task described in study~\cite{xu2024distinguishing}.
During the training phase, adversarial samples are generated from the original \datasetname{} training set using \attacktool{}, forming an augmented training set comprising both original and adversarial samples. 
This augmented dataset is utilized for adversarial fine-tuning of the model.
For performance evaluation, we first assess CodeGPTSensor and \toolname on the original \datasetname{} test set to validate their detection capabilities on unmodified code. 
Subsequently, to examine robustness against code perturbations, we generate an adversarial test set (\datasetname{}-adv) by applying \attacktool{} to modify samples in the original \datasetname{} test set. 
A comparative analysis of both tools on \datasetname{}-adv quantifies their robustness differences under adversarial scenarios.

\subsection{Selected Baselines}
\label{baselines}

\subsubsection{LLM-generated Code Detection}
Existing AIGC detectors can be divided into commercial ones~\cite{zerogpt,chatgptzero,copyleaks,writer,aitextclassifier} and open-source ones~\cite{gpt2detector,mitchell2023detectgpt, robertaqa}. 
These detectors are mainly designed to detect AI-generated natural language text, such as articles and essays, and their performance in detecting AI-generated code has not been comprehensively evaluated.
CodeGPTSensor is the first detector specially designed for detecting LLM-generated program code.
The experimental results presented in study~\cite{xu2024distinguishing} demonstrate that CodeGPTSensor achieves state-of-the-art performance in LLM-generated code detection tasks when analyzing unmodified code snippets. 
Building on this finding, we adopt CodeGPTSensor as the baseline method to rigorously evaluate the effectiveness of \toolname, an adversarially trained variant enhanced to address subtly modified adversarial samples. 
Our evaluation protocol specifically investigates whether \toolname can (1) preserve high detection accuracy on the original test set and (2) maintain robustness against adversarial perturbations introduced via minor code modifications, thereby comprehensively assessing its practical utility in real-world LLM-generated code detection scenarios.

\subsubsection{Adversarial Sample Generation}
\label{sec:attack-baselines}
To evaluate the effectiveness of \attacktool{}, this experiment compares it against three state-of-the-art black-box adversarial attack algorithms, described as follows:

\begin{itemize}[leftmargin=*]
\item \textbf{ALERT}~\cite{yang2022natural} generates adversarial samples through context-aware identifier prediction, employing a hybrid strategy that combines greedy search and genetic algorithms for identifier substitution.
\item \textbf{MOAA}~\cite{zhou2024evolutionary} utilizes context-aware identifier prediction to generate adversarial samples, leveraging a multi-objective genetic algorithm to balance the attack success rate and the extent of identifier modifications.
\item \textbf{CODA}~\cite{tian2023code} proposes the use of reference samples to narrow the search space for adversarial sample generation, first generating candidate samples via structural transformations and then performing identifier substitution on the optimal candidates.
\end{itemize}

\subsection{Evaluation Metrics}
\label{sec:eval_metrics}

\subsubsection{LLM-generated Code Detection}
To evaluate the effectiveness of \toolname in detecting LLM-generated code, we adopt the following evaluation metrics: Accuracy, Precision, Recall, F1-score, and AUC, which are widely used in the literature~\cite{ni2022best,ni2020jitjs,fu2022vulrepair}.

For each function, there are four possible detection results:
it can be detected as LLM-generated if it is exactly generated by LLM (true positive, TP); 
it can be detected as LLM-generated while it is human-written (false positive, FP); 
it can be detected as human-written while it is LLM-generated (false negative, FN); 
or it can be detected as human-written when it is exactly human-written (true negative, TN).
Therefore, based on the four possible results, evaluation metrics can be defined as follows:

\textbf{Accuracy} evaluates the performance of how many functions can be correctly classified.
It is calculated as $\frac{TP+ TN}{TP+FP+TN +FN}$.

\textbf{Precision} is the fraction of LLM-generated functions among the detected positive instances, which can be calculated as $\frac{TP}{TP+FP}$.

\textbf{Recall} measures how many LLM-generated functions can be correctly detected, which is defined as $\frac{TP}{TP+FN}$.

\textbf{F1-score} is a harmonic mean of $Precision$ and $Recall$, which can be calculated as $\frac{2 \times P \times R}{P + R}.$

\textbf{AUC} is the area under the receiver operating characteristic (ROC) curve, which is a 2D illustration of true positive rate (TPR) on the y-axis versus false positive rate (FPR) on the x-axis. 
ROC curve can be obtained by varying the classification threshold over all possible values.
The AUC score ranges from 0 to 1, and a well-performing classifier provides a value close to 1.

\subsubsection{Adversarial Sample Generation}
\label{sec:attack-metrics}
This experiment calculates evaluation metrics on a subset $X$ of the \datasetname{} test set, where each code snippet $x \in X$ satisfies the following conditions: (1) $x$ can be successfully parsed by Tree-sitter, and (2) $x$ contains at least one identifier, and (3) $x$ is correctly predicted by the target model $M$, i.e., $M(x)$ matches the ground truth label. 
Under these conditions, for a given $x$, if an adversarial sample $x^\text{adv}$ generated by an attack method causes $M$ to produce an incorrect prediction, $x^\text{adv}$ is considered a \textbf{successful adversarial sample}. 
To evaluate the adversarial attack performance of \attacktool{} and baseline algorithms, this experiment adopts three widely used metrics in adversarial attack research for deep code models~\cite{yang2022natural, zhou2024evolutionary, tian2023code, du2023extensive}:

\begin{itemize}[leftmargin=*]
\item \textbf{Attack Success Rate (ASR).} ASR measures the proportion of successful adversarial samples relative to the total number of test samples, defined as:

\begin{equation}
\frac{|{x|x\in X\wedge M(x) \neq M(x^{\text{adv}})}|}{|X|},
\end{equation}
\vspace{0.1cm}

a higher ASR indicates greater attack effectiveness.

\item \textbf{Average Model Queries (AMQ)}. 
In black-box adversarial attack scenarios, target models are often deployed on remote servers, and frequent queries can incur significant costs or risk service limitations. 
Following related studies~\cite{zhou2024evolutionary, du2023extensive}, this experiment uses AMQ to measure the average number of queries required to generate the first successful adversarial sample.

\item \textbf{Identifier Change Rate (ICR)}. 
ICR measures the proportion of identifiers replaced in successful adversarial samples relative to the total number of identifiers. 
A lower ICR indicates fewer identifier changes are needed to generate successful adversarial samples.
\end{itemize}

Additionally, to compare the quality of adversarial samples generated by different algorithms, this experiment calculates the \textbf{Semantic Distance (SD)} metric (measuring semantic consistency of adversarial samples, as defined in \autoref{equ:semantic-distance}) and the \textbf{Edit Distance (ED)} metric (measuring character-level differences between $x$ and $x^\text{adv}$, i.e., the number of insertions, deletions, and substitutions required to transform $x$ into $x^\text{adv}$, as defined in \autoref{equ:edit-distance}), aligning with the optimization objectives proposed in \autoref{sec:attack-module}.

In summary, an ideal adversarial attack algorithm should excel across all metrics, achieving a higher ASR while minimizing AMQ, ICR, SD, and ED.

\subsection{Implementation Details}
\label{implement_details}
Both GPTCodeSensor and \toolname models adopt a classification threshold of 0.5.
The baseline GPTCodeSensor model refers to the optimal model trained under the hyperparameters described in study~\cite{xu2024distinguishing}. 
For \toolname, adversarial fine-tuning retains identical hyperparameters (batch size=8, learning rate=2e-5, Adam optimizer) as specified in study~\cite{xu2024distinguishing}, but limits training to 1 epoch.  
The \attacktool{} module in \toolname follows Zhou et al.~\cite{zhou2024evolutionary} with population size $N=30$ and maximum iterations $\textit{max\_iter}=5 \times N_{var}$, where $N_{var}$ denotes the count of deduplicated identifiers in code snippets. 
For mutation operations, we set the identifier replacement probability $r=0.5$ and configure CodeT5 to generate $k=40$ candidate identifiers per iteration.
All these models are implemented in Python using the PyTorch framework.
The evaluation is conducted on a 16-core workstation equipped with an Intel(R) Xeon(R) Gold 6226R CPU @ 2.90Ghz, 192GB RAM, and 10 × NVIDIA RTX 3090 GPU, running Ubuntu 20.04.1 LTS.

\section{Experiment Results}
\label{sec:results}

To investigate the effectiveness of \toolname and MIST, our experiments focus on the following three research questions:

\begin{itemize}[leftmargin=*]
\item \textit{RQ-1 Detection Performance Comparison: \toolname vs. CodeGPTSensor.}

\item \textit{RQ-2 Adversarial Attack Performance Comparison: MIST module in \toolname vs. Baselines.} 

\item \textit{RQ-3 Robustness Enhancement via Adversarial Training: MIST-generated vs. Baseline-generated Adversarial Samples.}

\end{itemize}

\subsection{\bf{[RQ-1]: Detection Performance Comparison: \toolname vs. CodeGPTSensor.}} 
\label{sec:rq1}

\noindent
\textbf{Objective.}
In study~\cite{xu2024distinguishing}, CodeGPTSensor demonstrated exceptional LLM-generated code detection performance on the HMCorp dataset.
However, in real-world scenarios where LLM-generated code undergoes modifications, its detection accuracy may degrade significantly. 
To address this limitation and enhance robustness, this study proposes \toolname, a detection framework fortified via adversarial training to improve model resilience against input perturbations. 
Our empirical evaluation evaluates both \toolname and CodeGPTSensor on the original HMCorp test set and an adversarial variant (HMCorp-adv) containing subtly modified samples, verifying whether \toolname achieves dual objectives: maintaining high detection accuracy on unmodified samples while enhancing robustness against adversarially perturbed code.

\begin{table}[htbp]
  \centering
  \caption{\label{tab:rq1}Performance evaluation results of \toolname and CodeGPTSensor models on the original test set and adversarial test set}
  \resizebox{\linewidth}{!}{
    \begin{tabular}{llccccc}
    \toprule
    \textbf{Test Set} & \textbf{Model} & \multicolumn{1}{c}{\textbf{Accuracy}} & \multicolumn{1}{c}{\textbf{Recall}} & \multicolumn{1}{c}{\textbf{Precision}} & \multicolumn{1}{c}{\textbf{F1}} & \multicolumn{1}{c}{\textbf{AUC}} \\
    \midrule
    \multirow{3}[1]{*}{HMCorp-Java} & CodeGPTSensor & 0.967  & \textbf{0.976} & 0.959  & 0.968  & 0.995  \\
          & CodeGPTSensor+ & \textbf{0.970} & 0.973  & \textbf{0.967} & \textbf{0.970} & \textbf{0.996} \\
          \rowcolor{lightgray} \cellcolor{white} & \textit{Improvement} & \textit{0.3\%} & \textit{-0.4\%} & \textit{0.8\%} & \textit{0.2\%} & \textit{0.1\%} \\
    \midrule
    \multirow{3}[1]{*}{HMCorp-Python} & CodeGPTSensor & 0.992  & 0.991  & \textbf{0.993} & 0.992  & 0.999  \\
          & CodeGPTSensor+ & \textbf{0.992} & \textbf{0.992} & 0.992  & \textbf{0.992} & \textbf{0.999} \\
          \rowcolor{lightgray} \cellcolor{white} & \textit{Improvement} & \textit{0.0\%} & \textit{0.1\%} & \textit{0.0\%} & \textit{0.0\%} & \textit{0.1\%} \\
    \midrule
    \multirow{3}[1]{*}{HMCorp-Java-adv} & CodeGPTSensor & 0.325  & 0.065  & 0.785  & 0.120  & 0.822  \\
          & CodeGPTSensor+ & \textbf{0.944} & \textbf{0.974} & \textbf{0.949} & \textbf{0.961} & \textbf{0.983} \\
          \rowcolor{lightgray} \cellcolor{white} & \textit{Improvement} & \textit{190.8\%} & \textit{1396.8\%} & \textit{20.8\%} & \textit{699.6\%} & \textit{19.7\%} \\
    \midrule
    \multirow{3}[1]{*}{HMCorp-Python-adv} & CodeGPTSensor & 0.260  & 0.694  & 0.293  & 0.412  & 0.140  \\
          & CodeGPTSensor+ & \textbf{0.969} & \textbf{0.979} & \textbf{0.941} & \textbf{0.960} & \textbf{0.995} \\
          \rowcolor{lightgray} \cellcolor{white} & \textit{Improvement} & \textit{272.8\%} & \textit{41.0\%} & \textit{221.0\%} & \textit{132.8\%} & \textit{608.9\%} \\
    \bottomrule
    \end{tabular}%
  }
\end{table}%

\noindent
\textbf{Experimental Design.}
This experiment employs the implementation details described in Section~\ref{implement_details}, selecting CodeGPTSensor as the baseline model for comparative analysis. 
We adopt accuracy, precision, recall, F1-score, and AUC as comprehensive evaluation metrics. 
Performance evaluations are conducted on both the original HMCorp test set and the adversarial variant HMCorp-adv, which contains samples generated by the MIST module to mimic minor code modifications encountered in real-world scenarios. 
This dual-testbed approach rigorously assesses whether \toolname maintains its detection efficacy on unperturbed inputs while demonstrating enhanced robustness against adversarial perturbations compared to the baseline model.

\noindent 
\textbf{Results.}
The experimental results, as shown in Table~\ref{tab:rq1}, include an \textit{``Improvement''} row that quantifies the relative performance gains of \toolname over CodeGPTSensor across all metrics. 
On the original test sets, \toolname achieves detection accuracy of 0.970 and 0.992 on HMCorp-Java and HMCorp-Python, respectively, demonstrating minimal performance divergence from CodeGPTSensor, with nearly identical metrics across the board.
In contrast, \toolname significantly outperforms CodeGPTSensor on the adversarial test sets, exhibiting substantial improvements across all evaluation metrics. 
Specifically, on HMCorp-Java-adv, the detection accuracy increases from 0.325 to 0.944, while on HMCorp-Python-adv, it rises from 0.260 to 0.969.

\vspace{0.1cm}
\intuition{
\textbf{Answer to RQ-1}: 
\toolname not only maintains high detection accuracy on unmodified samples but also significantly enhances accuracy on adversarially perturbed samples, offering a more robust solution for LLM-generated code detection in real-world scenarios.
}

\subsection{\bf{[RQ-2]: Adversarial Attack Performance Comparison: MIST module in \toolname vs. Baselines.}}
\label{sec:rq2}

\noindent
\textbf{Objective.}
This experiment aims to compare the performance of MIST, the adversarial sample generation algorithm integrated into \toolname{}, with state-of-the-art black-box adversarial sample generation algorithms in adversarial attack tasks.
The quality of adversarial samples is critical for effective adversarial training. 
However, existing baseline algorithms either rely solely on identifier substitution or treat identifier substitution and structural transformations as isolated processes, neglecting the human tendency to alternate between variable renaming and structural adjustments during code editing. 
This limitation reduces the diversity and effectiveness of adversarial samples, thereby constraining the robustness improvements achieved through adversarial training. 
Furthermore, some baseline algorithms prioritize the attack success rate as the sole optimization objective, potentially resulting in excessive perturbations.
To address these issues and enhance the quality of adversarial samples, this study proposes MIST, a multi-objective optimization-based adversarial sample generation module that integrates identifier substitution strategies with structural transformation strategies.
MIST employs a multi-objective optimization framework to balance the attack success rate, semantic consistency, and perturbation magnitude. 
This experiment evaluates MIST against baseline algorithms to validate its superiority in both attack success rate and adversarial sample quality.

\noindent
\textbf{Experimental Design.}
We present the baseline algorithms, target models, datasets, and evaluation metrics employed in the adversarial attack experiments.

\textbf{\underline{(1) Baselines:}} To evaluate the effectiveness of \attacktool{}, this experiment compares it against three SOTA black-box adversarial attack algorithms, as described in Section~\ref{sec:attack-baselines}.
Specifically, for ALERT, MOAA, and CODA, this experiment adopts their official open-source implementations from GitHub~\cite{gitALERT, gitMOAA, gitCODA}, with necessary modifications to their dataset loading, target model integration, and invocation components to ensure compatibility with the \taskone{} model. 
For CODA~\cite{tian2023code}, parameters such as the number of reference samples and candidate samples generated after structural transformations follow the default settings described in its paper and implementation. 
For ALERT~\cite{yang2022natural}, MOAA~\cite{zhou2024evolutionary}, and \attacktool{}, this experiment aligns with Zhou et al.~\cite{zhou2024evolutionary}, setting the population size N to 30 and the maximum iteration count \textit{max\_iter} to $5 \times N_{var}$, where $N_{var}$ represents the number of unique identifiers in the sample. 
The remaining parameters for ALERT's greedy and genetic algorithms adhere to the defaults specified in its paper and implementation.
For \attacktool{}'s mutation operations, the identifier substitution probability r is set to 0.5, and the number of candidate identifiers generated using CodeT5, k, is set to 40, consistent with MOAA.

\textbf{\underline{(2) Target Models:}} To validate the effectiveness of \attacktool{} in adversarial attacks against \taskone{} models, this experiment selects CodeGPTSensor, the best-performing model for \taskone{}, and GPTSniffer, the state-of-the-art baseline model, as the target models. 
\taskone{} is a classification task aimed at determining whether a given code snippet is human-written or LLM-generated. 
Correspondingly, the objective of adversarial attacks is to mislead the classification model into misjudging the origin of the code.

\textbf{\underline{(3) Datasets:}} This experiment leverages the \datasetname{} dataset introduced in study~\cite{xu2024distinguishing}. 
All target models have been thoroughly fine-tuned on the \datasetname{} training set. 
The performance of adversarial attack algorithms is evaluated on the \datasetname{} test set, which includes code samples in both Java and Python programming languages.

\textbf{\underline{(4) Evaluation Metrics:}} This experiment leverages the metrics described in Section~\ref{sec:attack-metrics}. 
An ideal adversarial attack algorithm should excel across all metrics, achieving a higher ASR while minimizing AMQ, ICR, SD, and ED.
Furthermore, to quantitatively compare the overall performance of different adversarial attack methods, this experiment adopts the \textbf{TOPSIS-based comprehensive evaluation method}~\cite{TopsisPy}.
TOPSIS (Technique for Order Preference by Similarity to Ideal Solution) is a widely used multi-criteria decision analysis approach that calculates a composite score for each method based on its distance to the positive ideal solution (PIS) and negative ideal solution (NIS). 
Specifically, ASR is treated as a positive indicator (higher values are preferable), while the remaining metrics are treated as negative indicators (lower values are preferable).
Two sets of weights are applied in the TOPSIS calculation: (1) an equal-weight configuration, where each metric is assigned a weight of 0.2, to evaluate balanced performance across all metrics; 
and (2) an ASR-prioritized configuration, where ASR is assigned a weight of 0.6, and the remaining metrics are assigned weights of 0.1 each, to assess performance with a focus on attack success rate. 
Based on these weight configurations, composite scores are calculated and ranked for each adversarial attack method, providing an intuitive comparison of their overall effectiveness.

\noindent 
\textbf{Results.}
\autoref{tab:attack_results} presents the performance evaluation results of different adversarial attack methods against the \taskone{} model, with the best results for each metric highlighted in bold.
The analysis is conducted from three perspectives: attack success rate, adversarial sample quality, and attack efficiency.

\begin{table*}[htbp]
  \centering
  \caption{\label{tab:attack_results}Performance evaluation results of different adversarial attack methods on different LLM-generated detection models}
  \resizebox{\linewidth}{!}{
    \begin{tabular}{lllrrrrr}
    \toprule
    \textbf{Test Set} & \multicolumn{1}{l}{\textbf{Target Model}} & \textbf{Attack Method} & \textbf{ASR (\%)} & \textbf{ICR (\%)} & \textbf{SD} & \textbf{ED} & \textbf{AMQ} \\
    \midrule
    \multirow{8}[3]{*}{HMCorp-Java} & \multicolumn{1}{l}{\multirow{4}[2]{*}{GPTSniffer~\cite{NGUYEN2024112059}}} & ALERT~\cite{yang2022natural} & 18.03  & 58.26  & \textbf{0.70} & \textbf{46.40} & 240.60  \\
          &       & CODA~\cite{tian2023code}  & 30.58  & 90.13  & 1.69  & 94.88  & 348.35  \\
          &       & MOAA~\cite{zhou2024evolutionary}  & \textbf{59.94} & 36.08  & 1.95  & 145.29  & 243.20  \\
          &  & \cellcolor{lightgray}MIST  &  \cellcolor{lightgray} 48.41  &  \cellcolor{lightgray} \textbf{17.85} &  \cellcolor{lightgray} 0.89  &  \cellcolor{lightgray} 92.82  &  \cellcolor{lightgray} \textbf{97.29} \\
    \cmidrule{2-8} & \multicolumn{1}{l}{\multirow{4}[1]{*}{CodeGPTSensor~\cite{xu2024distinguishing}}} & ALERT~\cite{yang2022natural} & 20.62  & 62.61  & 0.69  & \textbf{39.43} & 184.46  \\
          &       & CODA~\cite{tian2023code}  & 29.36  & 88.62  & 1.69  & 95.05  & 347.98  \\
          &       & MOAA~\cite{zhou2024evolutionary}  & 49.62  & 26.22  & 1.43  & 132.45  & 195.23  \\
         \rowcolor{lightgray} \cellcolor{white} & \cellcolor{white} & MIST & \textbf{53.80} & \textbf{9.25} & \textbf{0.53} & 94.68  & \textbf{58.87} \\
    \midrule
    \multirow{8}[3]{*}{HMCorp-Python} & \multicolumn{1}{l}{\multirow{4}[1]{*}{GPTSniffer~\cite{NGUYEN2024112059}}} & ALERT~\cite{yang2022natural} & 6.57  & 62.56  & 0.60  & \textbf{44.78} & 243.55  \\
          &       & CODA~\cite{tian2023code}  & 7.62  & 85.80  & 1.62  & 70.49  & 348.11  \\
          &       & MOAA~\cite{zhou2024evolutionary}  & 51.95  & \textbf{7.25} & 0.53  & 281.19  & \textbf{14.53} \\
          &  & \cellcolor{lightgray}MIST &  \cellcolor{lightgray} \textbf{74.06} &  \cellcolor{lightgray} 7.42  &  \cellcolor{lightgray} \textbf{0.46} &  \cellcolor{lightgray} 139.11  &  \cellcolor{lightgray} 61.95  \\
    \cmidrule{2-8}  & \multicolumn{1}{l}{\multirow{4}[2]{*}{CodeGPTSensor~\cite{xu2024distinguishing}}} & ALERT~\cite{yang2022natural} & 6.22  & 50.89  & 0.51  & \textbf{30.88} & 167.47  \\
          &       & CODA~\cite{tian2023code}  & 6.76  & 76.56  & 1.45  & 61.92  & 341.99  \\
          &       & MOAA~\cite{zhou2024evolutionary}  & 53.55  & 8.96  & 0.63  & 279.12  & \textbf{41.66} \\
          \rowcolor{lightgray} \cellcolor{white} & \cellcolor{white} & MIST & \textbf{60.15} & \textbf{6.14} & \textbf{0.42} & 144.48  & 49.14  \\
    \bottomrule
    \end{tabular}%
  }
\end{table*}%

\textbf{(1) Attack Success Rate:} \attacktool{} demonstrates superior ASR, particularly on the Python test set, where it achieves ASRs of 74.06\% and 60.15\% against GPTSniffer and CodeGPTSensor, respectively, significantly outperforming other methods. On the Java test set, \attacktool{} achieves an ASR of 53.80\% against CodeGPTSensor, surpassing all baseline algorithms.

\textbf{(2) Adversarial Sample Quality:} \attacktool{} excels in ICR, particularly on the Java test set, where it achieves ICRs of 17.85\% and 9.25\% against GPTSniffer and CodeGPTSensor, respectively, which is far lower than other methods (baseline algorithms generally exceed 26\%). 
In terms of SD, \attacktool{} outperforms other methods in most cases, particularly on the Python test set, where it achieves the best SD for both target models. On the Java test set, \attacktool{} achieves an SD of 0.53 against CodeGPTSensor, significantly better than MOAA (1.43) and CODA (1.69). For ED, \attacktool{} shows a clear advantage, with ED values consistently lower than MOAA, which has the highest ASR among baseline algorithms. ALERT achieves the best ED but has the lowest ASR among all methods.

\textbf{(3) Attack Efficiency:} \attacktool{} also demonstrates advantages in AMQ, particularly on the Java test set, where it achieves AMQs of 97.29 and 58.87 against GPTSniffer and CodeGPTSensor, respectively, significantly lower than other methods (baseline algorithms generally exceed 184). On the Python test set, \attacktool{}'s AMQ is slightly higher than MOAA but remains significantly lower than ALERT and CODA.

\begin{figure*}[htbp]
    \centering
    \includegraphics[width=\linewidth]{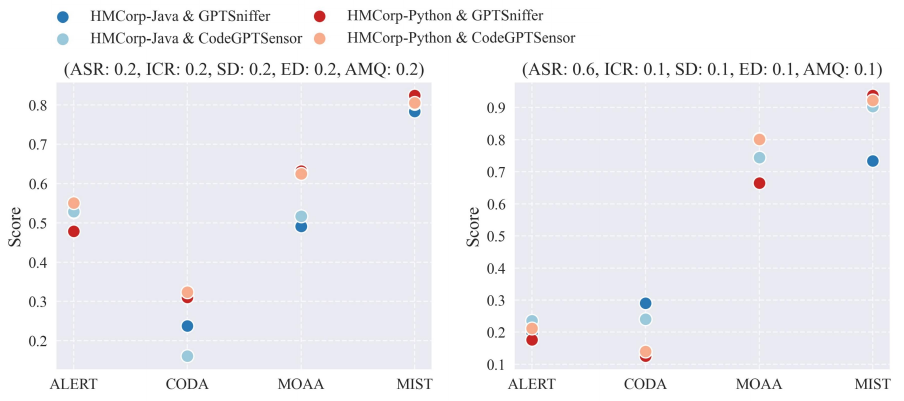}
    \caption{\label{fig:topsis}TOPSIS comprehensive score of different attack methods}
\end{figure*}

\autoref{fig:topsis} shows the composite scores of each adversarial attack method calculated using TOPSIS~\cite{TopsisPy} under two weight configurations. The results indicate that when equal weights are assigned to all five metrics (left figure), \attacktool{} achieves the highest composite score, significantly outperforming other methods.
When higher weight is assigned to ASR (right figure), \attacktool{} achieves the highest composite score in most cases, surpassing other methods.

In summary, \attacktool{} achieves a balanced performance in attack success rate, adversarial sample quality, and attack efficiency, highlighting its overall superiority in adversarial attacks against the \taskone{} model.
Therefore, in \toolname, this study employs the \attacktool{} module to generate adversarial samples, ensuring their challenge and quality.

\vspace{0.1cm}
\intuition{
\textbf{Answer to RQ-2}: 
Both individual metric analysis and TOPSIS-based comprehensive evaluation demonstrate that \attacktool{} achieves a balanced performance in attack success rate, adversarial sample quality, and attack efficiency, highlighting its overall superiority in adversarial attacks against the \taskone{} model.
}

\subsection{\bf{[RQ-3]: Robustness Enhancement via Adversarial Training: MIST-generated vs. Baseline-generated
Adversarial Samples.}}

\noindent
\textbf{Objective.}
This experiment aims to compare the effectiveness of adversarial samples generated by \attacktool{} in \toolname and baseline algorithms in enhancing the robustness of the \taskone{} model through adversarial training.
By conducting adversarial training using adversarial sample sets generated by different algorithms, we evaluate their respective contributions to improving the model's resilience against adversarial attacks. 
This comparison validates the superiority of adversarial samples generated by \attacktool{} in enhancing model robustness during adversarial training.

\noindent
\textbf{Experimental Design.}
To compare the effectiveness of different adversarial sample sets in enhancing model robustness, this experiment adopts the methodology proposed by Tian et al.~\cite{tian2023code}. 
The test set $S$ (specifically HMCorp-Java or HMCorp-Python) is divided into two equal subsets ($S_1$and $S_2$) for adversarial fine-tuning and evaluation, respectively, to prevent data leakage.
For each sample in $S_1$, adversarial samples are generated using ALERT, CODA, MOAA, and \attacktool{}.
If an attack is successful, the first successful adversarial sample is selected; if unsuccessful, the sample that minimizes the target model's prediction probability for the correct class is chosen. 
These adversarial sample sets are then used to fine-tune the target model through adversarial training.
For each sample in $S_2$, adversarial attacks are performed using ALERT, CODA, MOAA, and \attacktool{}, with successful adversarial samples forming the evaluation set. 
In other words, for samples in the evaluation set, the original target model (without adversarial fine-tuning) fails to predict their labels correctly, resulting in 0\% accuracy.
Subsequently, the detection accuracy of the fine-tuned models on the evaluation set is tested to analyze the effectiveness of different adversarial sample sets in enhancing model robustness through adversarial training.

\begin{figure*}[htbp]
    \centering
    \includegraphics[width=\linewidth]{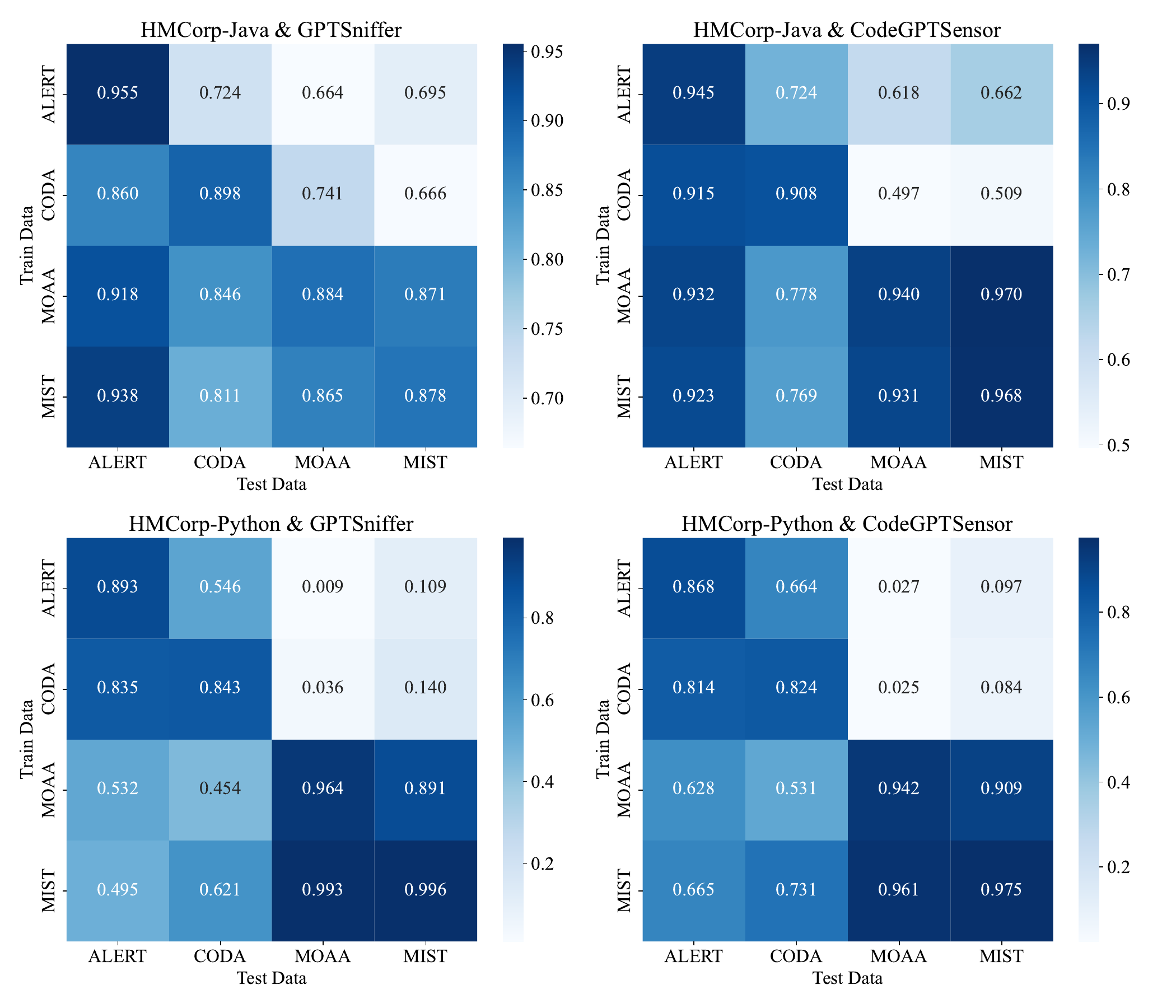}
    \caption{\label{fig:heatmap}The effect of adversarial samples generated by different adversarial attack algorithms on improving the robustness of the target model}
\end{figure*}

\noindent 
\textbf{Results.}
The experimental results, as shown in \autoref{fig:heatmap}, are presented in a heatmap format where the x-axis represents the source of the evaluation set, the y-axis represents the source of the data used for adversarial training, and the cell values indicate the detection accuracy of the fine-tuned target model under the corresponding conditions.
Based on the analysis of the results, the following conclusions can be drawn:

\textbf{(1) Models typically achieve optimal performance when the training and testing data sources are consistent:} The detection accuracy is generally highest when the training and testing data originate from the same adversarial attack method, indicating that the model gains defense capabilities specific to the attack method used during adversarial training.

\textbf{(2) Adversarial samples generated by multi-objective adversarial attack methods are more challenging:} The results show that models fine-tuned using adversarial samples generated by single-objective optimization methods (ALERT and CODA) exhibit lower detection accuracy when tested against adversarial samples generated by multi-objective methods (MOAA and \attacktool{}), as indicated by the light blue cells in the upper-right corner of the heatmap.

\textbf{(3) Adversarial samples generated by \attacktool{} effectively enhance model robustness:} When adversarial samples generated by \attacktool{} are used for adversarial training, the model achieves high detection accuracy across all test data sources. 
Notably, on the Python dataset, the detection accuracy of the target model against MOAA-generated test samples reaches 0.993 and 0.961 for GPTSniffer and \toolname{}, respectively, significantly higher than when other methods are used as training data sources. 
This demonstrates the superiority of \attacktool{} in generating adversarial samples that enhance the model's resilience to adversarial attacks.

In summary, the experimental results validate the superiority of \attacktool{}, a multi-objective optimization-based approach, in generating high-quality adversarial samples. 
The adversarial samples generated by \attacktool{} exhibit high attack success rates and low perturbation levels, posing greater challenges to the robustness of the target model. 
Furthermore, these samples effectively enhance the target model's resilience against diverse adversarial attacks during adversarial training.
Therefore, in \toolname, this study employs adversarial samples generated by \attacktool{} to construct the augmented training set, ensuring the effectiveness of adversarial training in improving model robustness and enabling the model to effectively defend against adversarial samples generated by various algorithms.

\vspace{0.1cm}
\intuition{
\textbf{Answer to RQ-3}: 
The experimental results demonstrate the superiority of adversarial samples generated by \attacktool{} in enhancing model robustness during adversarial training.
These samples simultaneously achieve high attack success rates and low perturbation levels, thereby posing greater challenges to the robustness of the target model. 
}

\section{Threats to Validity}

\noindent
\textbf{Threats to Internal Validity} primarily stem from potential errors in code implementation. 
To mitigate these threats, this study rigorously tested each module of \toolname during implementation to ensure functionality aligned with expectations. Additionally, the source code and experimental scripts were thoroughly reviewed multiple times to ensure the correctness of the implementation and the reliability of the experimental results.
Furthermore, for baseline algorithms, this study directly utilized the source code publicly available in the authors' GitHub repositories, adhering to the hyperparameter configurations provided in their respective papers, to ensure fairness and reliability in the experiments.

\noindent
\textbf{Threats to External Validity} primarily concern the generalizability and applicability of the experimental findings.
To address these threats, this study employed the large-scale \datasetname{} dataset for experiments, which includes code samples with strong diversity and representativeness. However, the study has the following limitations:
(1) The analysis is limited to Java and Python, excluding other widely used programming languages;
(2) The dataset is constructed based on a specific version of ChatGPT (gpt-3.5-turbo from April 2023), but future versions of LLMs may exhibit different code generation capabilities and styles. Thus, it is necessary to periodically evaluate \toolname's detection performance on code generated by newer LLMs and optimize the detection methods accordingly;
(3) The adversarial sample generation module \attacktool{} in \toolname primarily combines identifier substitution and structural transformation strategies to simulate minor code modifications in real-world scenarios. However, actual scenarios may involve more diverse code modifications, such as function splitting or the introduction of third-party libraries, which are beyond the perturbation strategies of \attacktool{}. 
Therefore, future work could expand its perturbation strategies and evaluate the extended method’s performance in more complex modification scenarios.

\noindent
\textbf{Threats to Construct Validity} mainly arise from the appropriateness of the selected performance metrics. 
To mitigate these threats, this study adopted widely used classification metrics, including accuracy, precision, recall, F1-score, and AUC, to comprehensively evaluate the \taskone{} performance of \toolname and CodeGPTSensor on both the original and adversarial test sets.

\label{sec:threats}

\section{Related Work}
\label{sec:related_work}

\subsection{LLM-generated Content Detection}
\subsubsection{LLM-generated Text Detection}
Large language models are undergoing continuous iterations, resulting in significantly improved performance on various language-related tasks and the ability to generate convincing text~\cite{choi2023chatgpt,zhang2022opt,mitchell2023detectgpt}.
The strong capabilities of ChatGPT have sparked considerable interest and concern.
For instance, individuals are either curious about how closely ChatGPT resembles human experts or apprehensive about the potential risks posed by LLMs like ChatGPT.
Therefore, to ensure the responsible and ethical use of LLM-generated content, it is necessary to propose approaches to identify whether a given piece of content is generated by LLMs.
Yang et al.~\cite{yang2023survey} categorize existing methods for detecting LLM-generated content into three groups: 1) Training-based methods~\cite{chatgptzero, aitextclassifier, zhan2023g3detector, chen2023gpt, yu2023gpt, liu2022coco, wu2023llmdet, tian2023multiscale, hu2024radar}, which typically involve fine-tuning a pre-trained language model on a constructed dataset containing both human-written and LLM-generated content; 
2) Zero-shot methods~\cite{mireshghallah2023smaller, krishna2024paraphrasing, mitchell2023detectgpt, yang2023dna}, which leverage intrinsic properties of typical LLMs, such as probability curves~\cite{mitchell2023detectgpt} and N-gram divergence~\cite{yang2023dna}, for self-detection; and 3) Watermarking, which involves embedding information within the generated text to facilitate the identification of its source.

Our proposed approach falls within the category of training-based methods. 
Unlike existing models that primarily focus on detecting text generated by LLMs using data sources such as Wikipedia~\cite{guo2023hc3} and student essays~\cite{verma2023ghostbuster}, our work specifically targets the identification of program code generated by LLMs.

\subsubsection{LLM-generated Code Detection}
In the domain of LLM-generated code detection, Pan et al.~\cite{pan2024assessing} conduct an empirical investigation utilizing a dataset derived from fundamental Python programming problems to assess the effectiveness of various AIGC detectors~\cite{chatgptzero, Sapling, gpt2detector, mitchell2023detectgpt, gehrmann2019gltr} in identifying AI-generated code.
Their results indicate that these detectors exhibit insufficient performance in recognizing AI-generated code, underscoring the pressing need for more robust detection methods, especially within educational settings.
The recent work by Nguyen et al.~\cite{NGUYEN2024112059} introduces a CodeBERT-based classifier called GPTSniffer.
In contrast to training-based methods, Yang et al.~\cite{yang2023zero} focus on the zero-shot detection of LLM-generated code and present DetectGPT4Code, a technique that employs a small proxy model to approximate the logits on the conditional probability curve. 
Additionally, drawing inspiration from earlier text watermarking techniques~\cite{kirchenbauer2023watermark}, Lee et al.~\cite{lee2023wrote} propose a watermarking approach for code generation. 
This method selectively injects watermarks into tokens with entropy exceeding a predefined threshold, aiming to enhance both the detectability and quality of the generated code.

However, in practical applications, LLM-generated code is often subjected to minor human modifications, such as variable renaming or code structure adjustments. 
While these modifications do not alter the core logic of the code, they can significantly impact the performance of existing detection models, leading to a sharp decline in detection accuracy.
Notably, even the SOTA model (i.e, CodeGPTSensor~\cite{xu2024distinguishing}) also demonstrates insufficient robustness when facing modified LLM-generated code.
To address these limitations, we propose \toolname{}, which uses adversarial training techniques to further enhance the robustness of the detection model when facing minor code modifications.

\subsection{Adversarial Attack Techniques for Code Pre-trained Models}
\label{sec:background-attack}

Recent years have witnessed growing research attention on adversarial attack techniques targeting code pre-trained models, driven by their expanding deployment in software engineering tasks. 
These models demonstrate state-of-the-art performance across various code intelligence tasks. 
However, empirical studies reveal that, akin to traditional deep learning models in computer vision and NLP domains, code pre-trained models exhibit vulnerability to adversarial perturbations~\cite{du2023extensive, henkel2022semantic, li2022ropgen, pour2021search, rabin2021generalizability, yang2022natural, zeng2022extensive, zhang2020generating, zhang2020adversarial}. 
Adversarial examples crafted through semantics-preserving code transformations can systematically deceive model predictions, posing critical security risks in practical applications. 
For instance, attackers could generate adversarial variants from vulnerability-laden code snippets that preserve exploitability while evading detection by vulnerability identification models.

Unlike continuous-space perturbations in computer vision, code-oriented adversarial attacks operate in discrete spaces while adhering to syntactic and semantic constraints. 
Furthermore, adversarial code transformations must preserve functional equivalence and grammatical correctness~\cite{yang2022natural, tian2023code}, introducing unique technical challenges absent in other domains. 
In the field of code, existing adversarial example generation methods are mainly divided into two categories: white-box attacks and black-box attacks.
White-box attack methods~\cite{yefet2020adversarial,srikant2021generating,zhang2022towards} rely on the internal information of the target model (e.g., gradients) to generate adversarial examples in a targeted manner.
While white-box attack methods demonstrate strong attack efficacy, their dependence on gradient-based information from target models' internal architectures imposes practical constraints in real-world deployment scenarios. 
This limitation becomes particularly pronounced when attacking prevalent closed-source models like ChatGPT, which are deployed via API-based services that strictly prohibit access to model parameters or gradient computations, thereby rendering white-box attack strategies fundamentally inapplicable.
Black-box attack methods operate under the assumption that adversaries cannot access the target model's internal architecture, parameters, or gradient computations, restricting their knowledge to observable input-output relationships obtained through iterative query interactions.
However, existing black-box attack methods~\cite{yang2022natural, tian2023code, zhou2024evolutionary} have limitations when generating adversarial examples.
On the one hand, most existing methods rely on identifier replacement strategies and fail to integrate code structure transformations effectively
This makes it difficult to simulate the behavior of humans who often alternate between variable renaming and logic restructuring when rewriting code. 
This limits the diversity and representativeness of adversarial examples. 
On the other hand, existing methods struggle to achieve an ideal balance between attack success rate, semantic consistency, and the degree of disturbance, resulting in adversarial examples that are either not challenging enough or deviate too far from the original examples.

Unlike prior methods, we propose an adversarial sample generation module called \textbf{M}ulti-objective \textbf{I}dentifier and \textbf{S}tructure \textbf{T}ransformation (MIST), which integrates identifier replacement and code structure transformation to simulate manual code modification. 
Through a multi-objective optimization framework, we can strike a balance between attack success rate, semantic consistency, and the degree of disturbance, thereby generating high-quality adversarial samples that are more representative and challenging.

\section{Conclusion}
To address the real-world scenario where LLM-generated code may undergo minor modifications, we propose \toolname, an enhanced version of CodeGPTSensor, which employs adversarial training to improve robustness against input perturbations. 
The proposed method integrates a multi-objective optimization-based adversarial sample generation module, \attacktool{}, to generate high-quality and representative adversarial samples, thereby effectively enhancing the model's resilience against diverse adversarial attacks.

Experimental results on the \datasetname{} dataset demonstrate that \toolname significantly improves detection accuracy on adversarially perturbed code snippets while maintaining high accuracy on the original test set, showcasing superior robustness compared to CodeGPTSensor. 
Evaluation results of \attacktool{} indicate that the generated adversarial samples achieve high attack success rates while maintaining low perturbation levels and computational overhead, outperforming baseline adversarial sample generation algorithms in overall performance. 
Furthermore, adversarial samples generated by \attacktool{} effectively enhance the detection model's ability to defend against multiple adversarial attacks during adversarial training.
\label{sec:conclusion}

\begin{acks}
This work was supported by the National Natural Science Foundation of China (Grant No.62202419), the Fundamental Research Funds for the Central Universities (No. 226-2022-00064),
Zhejiang Provincial Natural Science Foundation of China (No. LY24F020008),
the Ningbo Natural Science Foundation (No. 2022J184), 
and the State Street Zhejiang University Technology Center.
\end{acks}

\bibliographystyle{ACM-Reference-Format}
\bibliography{main}


\end{document}